\documentclass[twocolumn,floatfix,superscriptaddress,a4paper,showpacs,showkeys,nofootinbib,notitlepage]{revtex4-1}
\usepackage[colorlinks=true,linktocpage=true,linkcolor=blue,citecolor=blue,allcolors=blue]{hyperref}
\usepackage{epsfig}
\usepackage{latexsym}
\usepackage[utf8]{inputenc}
\usepackage{xspace}
\usepackage{indentfirst}
\usepackage{enumitem}
\usepackage{color}
\usepackage{placeins}

\usepackage{setspace}
\usepackage{lipsum}

\usepackage{hyperref}

\usepackage{todonotes}

\usepackage{amsmath}
\usepackage{amssymb}
\usepackage[english]{babel}
\usepackage{url}
\graphicspath{{figs/}}
\newcommand{\mean}[1]{\langle #1 \rangle}
\newcommand{\eq}[1]{\begin{align} #1 \end{align}}

\newcommand\ddfrac[2]{\frac{\displaystyle #1}{\displaystyle #2}}

\newcommand{\be}{\begin{equation}}
\newcommand{\ee}{\end{equation}}

\begin{document}
\title{
Possible origin of HADES data on proton number fluctuations in Au+Au collisions   
}
\author{Oleh Savchuk}
\affiliation{Bogolyubov Institute for Theoretical Physics, Kyiv, Ukraine}
\affiliation{
GSI Helmholtzzentrum f\"ur Schwerionenforschung GmbH, Planckstr. 1, D-64291 Darmstadt, Germany}
\affiliation{Frankfurt Institute for Advanced Studies, Giersch Science Center,
D-60438 Frankfurt am Main, Germany}

\author{Roman V. Poberezhnyuk}
   \affiliation{Bogolyubov Institute for Theoretical Physics, Kyiv, Ukraine}
  \affiliation{Frankfurt Institute for Advanced Studies, Giersch Science Center,
D-60438 Frankfurt am Main, Germany}

\author{Mark I. Gorenstein}
\affiliation{Bogolyubov Institute for Theoretical Physics, Kyiv, Ukraine}
\affiliation{Frankfurt Institute for Advanced Studies, Giersch Science Center,
D-60438 Frankfurt am Main, Germany}

\date{\today}
\begin{abstract}
Recent data of the HADES Collaboration in Au+Au central collisions at $\sqrt{s_{\rm NN}}=2.4$~GeV indicate large proton number fluctuations inside one unit of rapidity around midrapidity.
This 
can be a signature of critical phenomena due to the strong attractive interactions between baryons. 
We study an alternative 
hypothesis that these
large fluctuations 
are caused by the event-by-event fluctuations of the number of bare protons, and no interactions between these protons are assumed.
The proton number fluctuations 
in 
five symmetric rapidity intervals $\Delta y$ inside the region 
$\Delta Y=1$ are calculated using the binomial acceptance procedure. This procedure 
assumes  the independent (uncorrelated)  emission of protons, and it  appears to be  in agreement 
with the HADES data. To check this simple picture we suggest to calculate the correlation between proton multiplicities in non-overlapping rapidity intervals $\Delta y_1$ and $\Delta y_2$ placed inside $\Delta Y=1$.

\end{abstract}

\maketitle
\section{Introduction}

The investigation of the  phase diagram of strongly interacting matter is today
one of the important topics in nuclear and particle physics.
Transitions between different phases 
are expected to reveal themselves as specific patterns in particle number fluctuations. 
In particular, a critical point (CP)  should yield large fluctuations of the conserved charges 
\cite{Stephanov:1998dy,Stephanov:1999zu,Athanasiou:2010kw,Stephanov:2008qz,PhysRevC.86.024904,Vovchenko:2015uda}.
This generally applies not only to the hypothetical  QCD CP which has garnered most attention, but also to the better established CP of the nuclear liquid-gas transition ~\cite{Sauer:1976zzf,Pochodzalla:1995xy,10.2307/2980526,
Vovchenko:2015pya,
Bondorf:1995ua}.

The particle number fluctuations can be characterized by the central moments, $\langle(\Delta N)^{2}\rangle\equiv \sigma^2$, $\langle (\Delta N)^{3}\rangle$, $\langle (\Delta N)^{4}\rangle$, etc, where $\langle...\rangle$ denotes the event-by-event averaging and $\Delta N \equiv N -\langle N \rangle$. The scaled variance $\omega$, 
(normalized) skewness $S\sigma$, and  kurtosis $\kappa\sigma^{2}$ of particle number distribution are defined as the following combinations of the central moments,
\eq{ \omega[N] & =\frac{\sigma^2}{\mean{N}}~=~\frac{\kappa_2}{\kappa_1}~
,\label{omega}\\
S\sigma[N] & = \frac{\langle(\Delta N)^{3}\rangle}{\sigma^{2}}~=~\frac{\kappa_3}{\kappa_2}~, \label{skew}\\
 \kappa  \sigma^{2}[N] & = \frac{\langle(\Delta N)^{4}\rangle-3\langle(\Delta N)^{2}\rangle^2}{\sigma^{2}}~=~\frac{\kappa_4}{\kappa_2}~,\label{kurt}
 }
where $\kappa_n$ are 
the cumulants of the $N$-distribution.
The size-independent (intensive) measures of particle number fluctuations  (\ref{omega}-\ref{kurt}) 
are also applied to conserved charges such as net baryon number $B$ and electric charge $Q$.

Having generally longer equilibration times~\cite{Asakawa:2000wh,Jeon:2000wg}, the fluctuations of conserved charges are also thought to reflect properties of earlier stages of collision~\cite{PhysRevC.86.024904}.
Studies of the higher-order fluctuation measures are motivated by their larger sensitivity to critical phenomena
\cite{Stephanov:2008qz,Stephanov:2011pb, 
Schaefer:2011ex,Vovchenko:2015pya,Chen:2015dra,Poberezhnyuk:2019pxs}.
Experimental studies of such fluctuation measures are in progress~\cite{Luo:2017faz}.

Total baryon number and electric charge 
are conserved event-by-event.
Therefore, actual fluctuations of conserved charges can only be seen 
in finite acceptance regions. 
An optimal choice of acceptance 
is important problem.
If, on the other hand, acceptance is too small, the trivial Poisson-like
fluctuations dominate~\cite{Athanasiou:2010kw,PhysRevC.102.024908,PhysRevC.105.044903,PhysRevC.102.064906}.
The acceptance should be large enough compared to correlation lengths relevant for various physics processes
~\cite{Ling:2015yau,VOVCHENKO2020135868}.

The (net)baryon number fluctuations are expected to be an important signature of any critical phenomena. 
Because detecting neutrons is problematic, in practice the (net)proton number fluctuations are studied.   
In central nucleus-nucleus collisions the 
(net)proton fluctuations
are measured at different collision energies as a function of size of 
a rapidity interval $\Delta y$.
At high collision energies
fluctuations 
correspond to the Poisson distribution at small $\Delta y\ll 1$ and they  {\it decreases monotonously} with $\Delta y $.  An explanation of this behavior was recently considered in Refs.~\cite{PhysRevC.103.044903,SAVCHUK2022136983}. 
The main physical effects suppressing  the proton number fluctuations are the global baryon number conservation 
and excluded volume repulsive interactions between protons.  

Recently the HADES Collaboration data for proton number fluctuations~\cite{HADES:2020wpc} were reported for 5\% central Au+Au collisions at the center of mass collision energy of nucleon pairs $\sqrt{s_{\rm NN}}=2.42$~GeV. Note that at this small energy the antiproton production is negligible. In contrast to the data at RHIC and LHC energies the HADES results demonstrate 
that the scaled variance for protons {\it increases   monotonously}   with $\Delta y$ from unity at $\Delta y\ll 1$ to $\omega >2$ in the symmetric rapidity interval $\Delta Y=1$ in the center of mass system.
The effects of baryon conservation and repulsion that appear to drive the behavior of proton number cumulants at high energies fail to describe the HADES data even qualitatively~\cite{Vovchenko:2022syc}.

Large event-by-event fluctuations of proton number
can  potentially be a signal of abnormal hadron matter  equation of state. This possibility is discussed in Ref.~\cite{VOVCHENKO2022137368},
which requires strong correlations among the emitted protons in the coordinate space,
e.g. due to a possible presence of the critical point in the baryon-rich regime.
In the present paper we consider an alternative possibility when no interactions between the detected protons are assumed.

The paper is organized as follows. In Sec.~\ref{sec-binomial} we present the formulas of the 
binomial acceptance procedure 
which connect the fluctuations measures
in the finite acceptances with the corresponding quantities  
in the full phase space. In Sec.~\ref{HADES} the HADES results are fitted within the  the 
binomial acceptance  procedure.
Conclusions in Sec.~\ref{concl} closes the article.

\section{Binomial acceptance procedure}\label{sec-binomial}

To connect the fluctuations in different rapidity intervals 
we assume that acceptance of particles is binomial, i.e.
that each particle of a given type is accepted by detector with a fixed probability $\alpha$ \cite{PhysRevC.86.044904,PhysRevC.85.021901}. This probability $0\leq\alpha=\langle n\rangle/\langle N\rangle \leq 1$ equals the ratio of the mean number $\langle n\rangle$
of particles accepted in a fixed region of momentum space $\Delta y$ to the mean number $\langle N \rangle $  of  particles of the same type in a ``full" momentum space $\Delta Y$. A full momentum space does not necessary means complete $4\pi$-acceptance.  
The sufficient condition for $\Delta Y$ is to fully encompass $\Delta y$.
The main assumption of the binomial acceptance is that the probability $\alpha$ is the same for all particles of a given type and independent of any properties of a specific event. 
This assumption allows to relate the cumulants
within a finite acceptance 
to their values in the larger, encompassing phase space.

Let the function $P(N)$
be a normalized probability distribution for observing  $N$ particles of a given type in the ``full" phase space. 
Assuming the binomial acceptance for particles, the
probability $p(n,\alpha)$ to observe $n$ particles
detected
in the finite $\alpha$-region of the phase space is
\eq{\label{binomial}
p(n,\alpha)&=   
\sum_{N=n}^{\infty} \frac{N!}{n!(N-n)!}\,\alpha^n (1-\alpha)^{N-n}\,P(N)\nonumber\\&\equiv \sum_{N=n}^{\infty} B(N,n|\alpha)\,P(N)~.
}

The scaled variance, skewness, and kurtosis  of the accepted particles 
in $\Delta y\le 1$ 
are then presented  
using the distribution (\ref{binomial})
as follows \cite{PhysRevC.101.024917}:
\eq{
\label{w-alpha}
  \omega_{\alpha}[n]  & \equiv \frac{\kappa_2[n|\alpha]}{\kappa_1[n|\alpha]}= 1-\alpha+\alpha\omega[N]~,\\
\label{s-alpha}
 S\sigma_{\alpha}[n] & = \frac{\kappa_3[n|\alpha]}{\kappa_2[n|\alpha]}=\frac{\omega[N]}{\omega_{\alpha}[n]}\left\{\alpha^{2}S\sigma[N] + 3\alpha(1-\alpha) \right\}\nonumber \\& +\frac{1-\alpha}{\omega_{\alpha}[n]}(1-2\alpha)~,\\
\label{k-alpha}
\kappa\sigma^{2}_{\alpha}[n]  & = 
 \frac{\kappa_4[n|\alpha]}{\kappa_2[n|\alpha]}  =
    \ddfrac{\omega[N]}{\omega_{\alpha}[n]}\left\{\alpha^{3}\kappa\sigma^{2}[N] \right\}\\&+ \ddfrac{\omega[N]}{\omega_{\alpha}[n]}(1-\alpha)\left\{ 6\alpha^{2}S\sigma[N]+\alpha(7-11\alpha)\right\} \nonumber\\&+\ddfrac{1-\alpha}{\omega_{\alpha}[n]}\left\{1-6\alpha(1-\alpha)\right\}\nonumber~,
}
where $\omega$, $S\sigma$, and $\kappa \sigma^2$ correspond to acceptance interval $\Delta Y$ and
are given by Eqs.~(\ref{omega}-\ref{kurt}).

At $\alpha\to 1$ in Eqs.~(\ref{w-alpha}-\ref{k-alpha}), one evidently finds $\omega_{\alpha}[n]\to\omega[N]$, $S\sigma_{\alpha}[n]\to S\sigma[N]$, and $\kappa\sigma^2_{\alpha}[n]\to \kappa \sigma^2[N]$, i.e.,  the binomial acceptance results 
approach those in the full rapidity region $\Delta Y=1$.
In the opposite limit, $\alpha\to 0$, the cumulant ratios are Poissonian, and  $\omega_{\alpha}[n]$, $S\sigma_{\alpha}[n]$,  $\kappa\sigma^2_{\alpha}[n]\to 1$.

\begin{figure*}
\includegraphics[width=.37\textwidth]{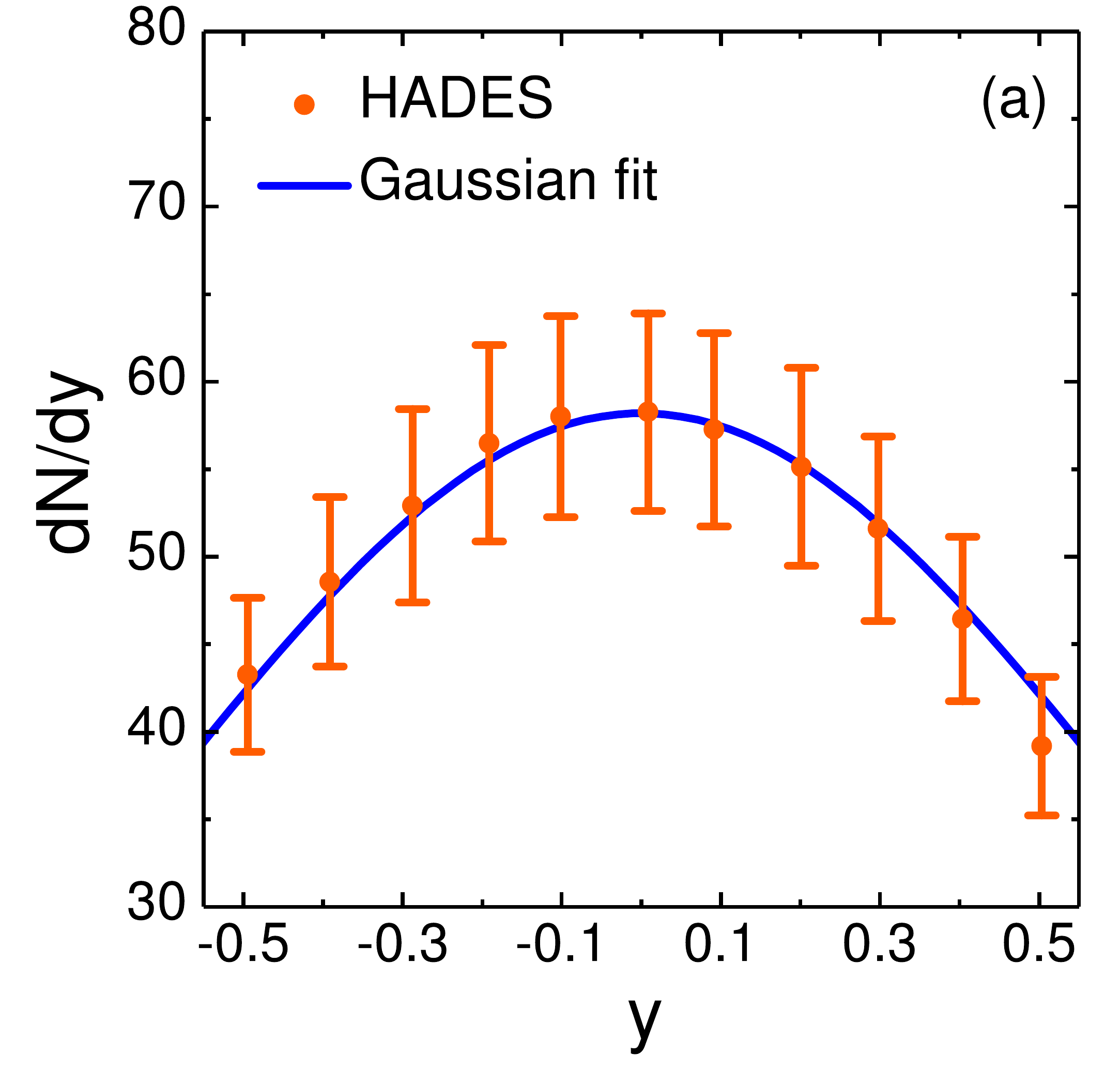}
\includegraphics[width=.37\textwidth]{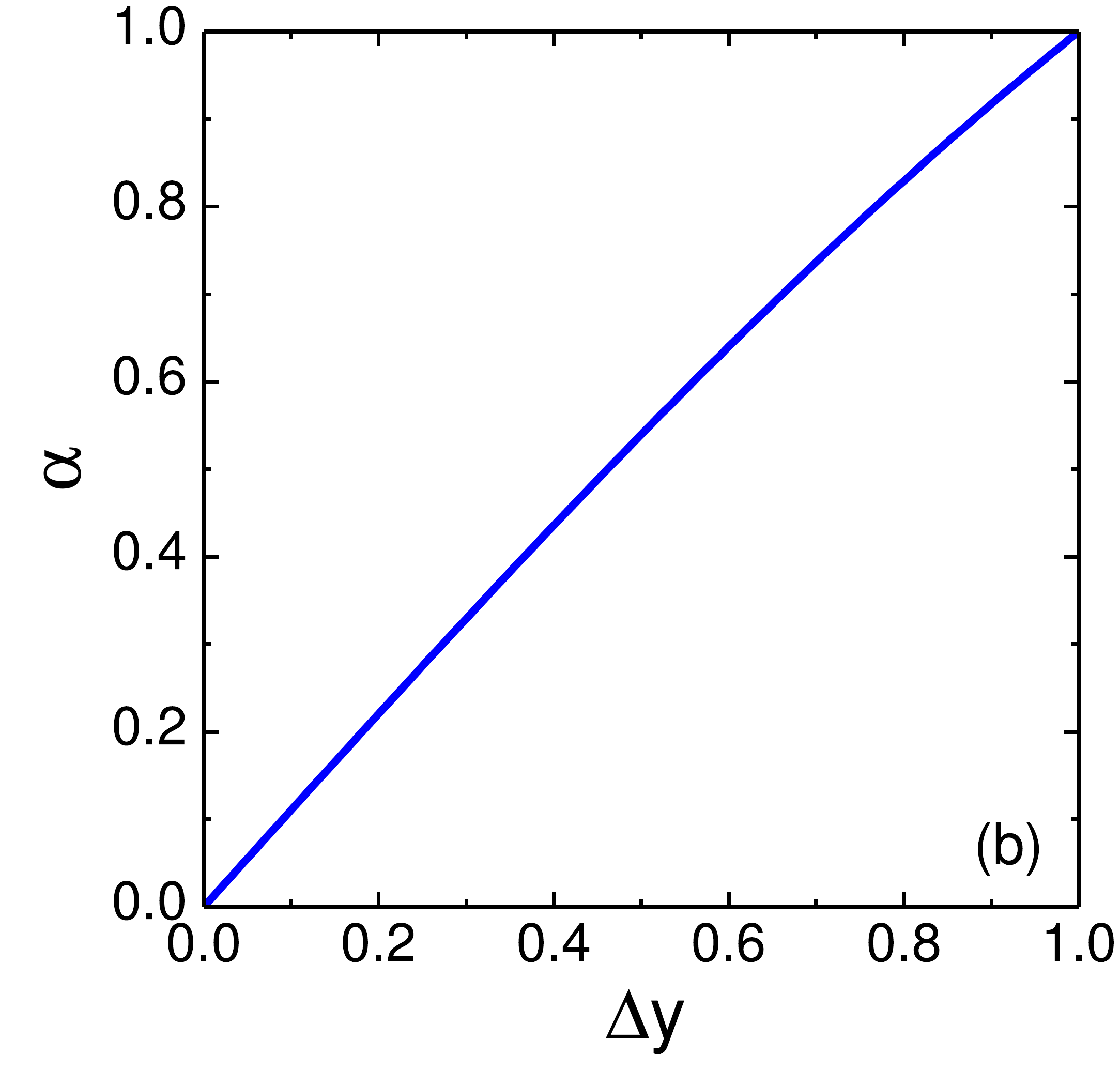}
\caption{(a): The HADES rapidity distribution  are presented by symbols and solid line shows the gaussian fit (\ref{gauss}). (b): Acceptance $\alpha$-parameter for the HADES data as a function of the rapidity interval $\Delta y$ calculated using Eq.~(\ref{x}).\label{fig}}
\end{figure*}
\begin{figure*}
\includegraphics[width=.32\textwidth]
{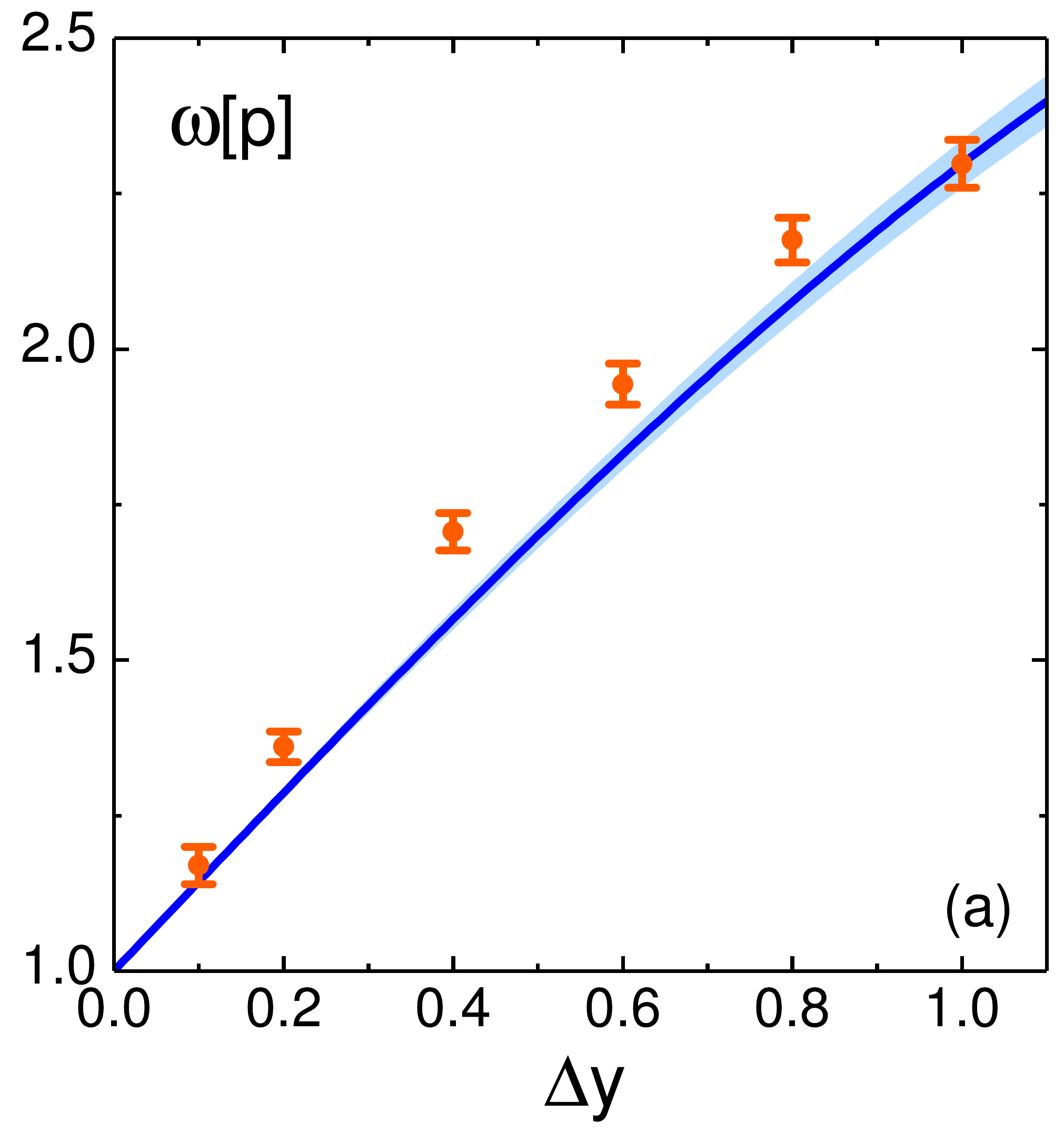}
\includegraphics[width=.32\textwidth]{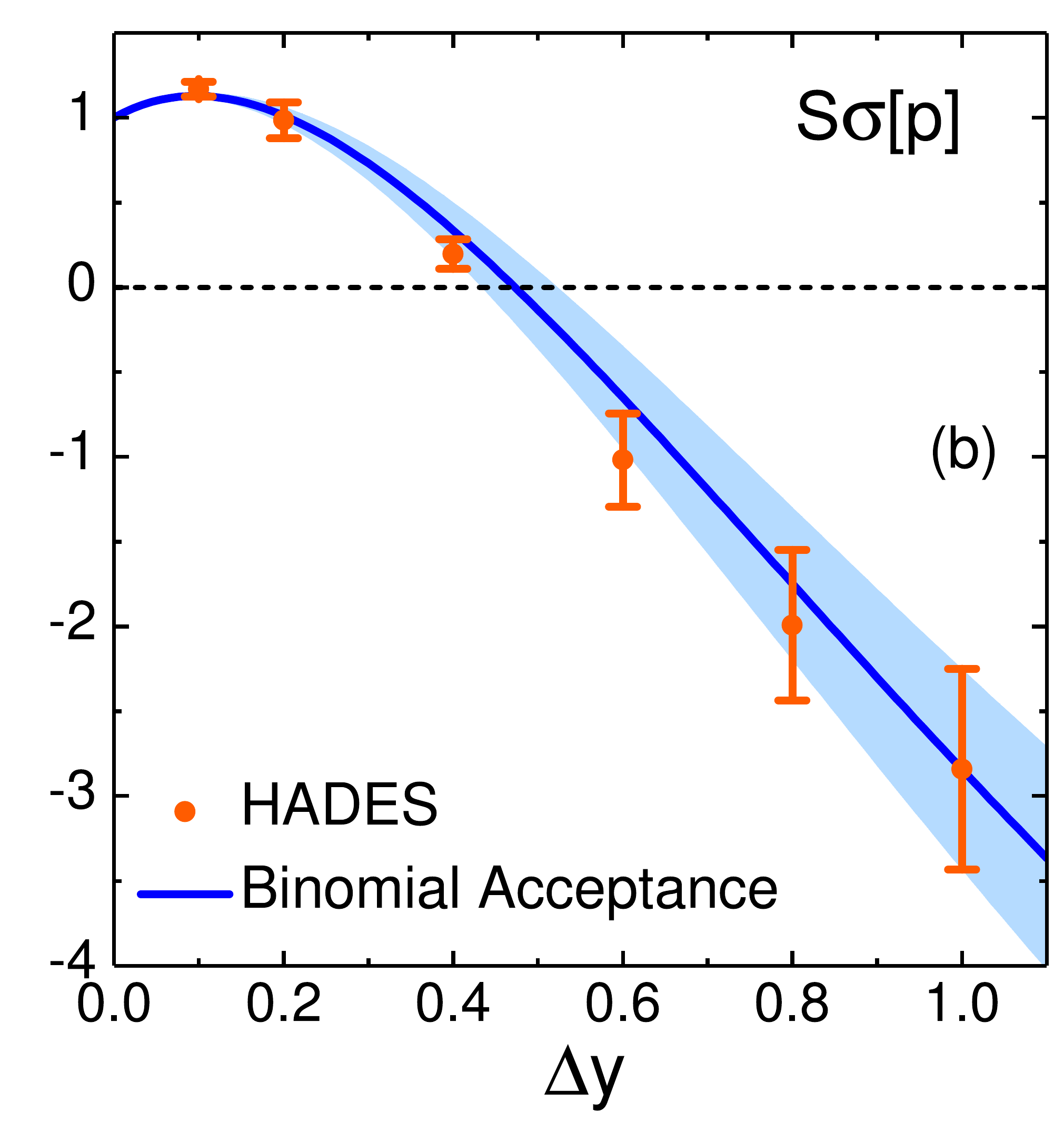}
\includegraphics[width=.32\textwidth]{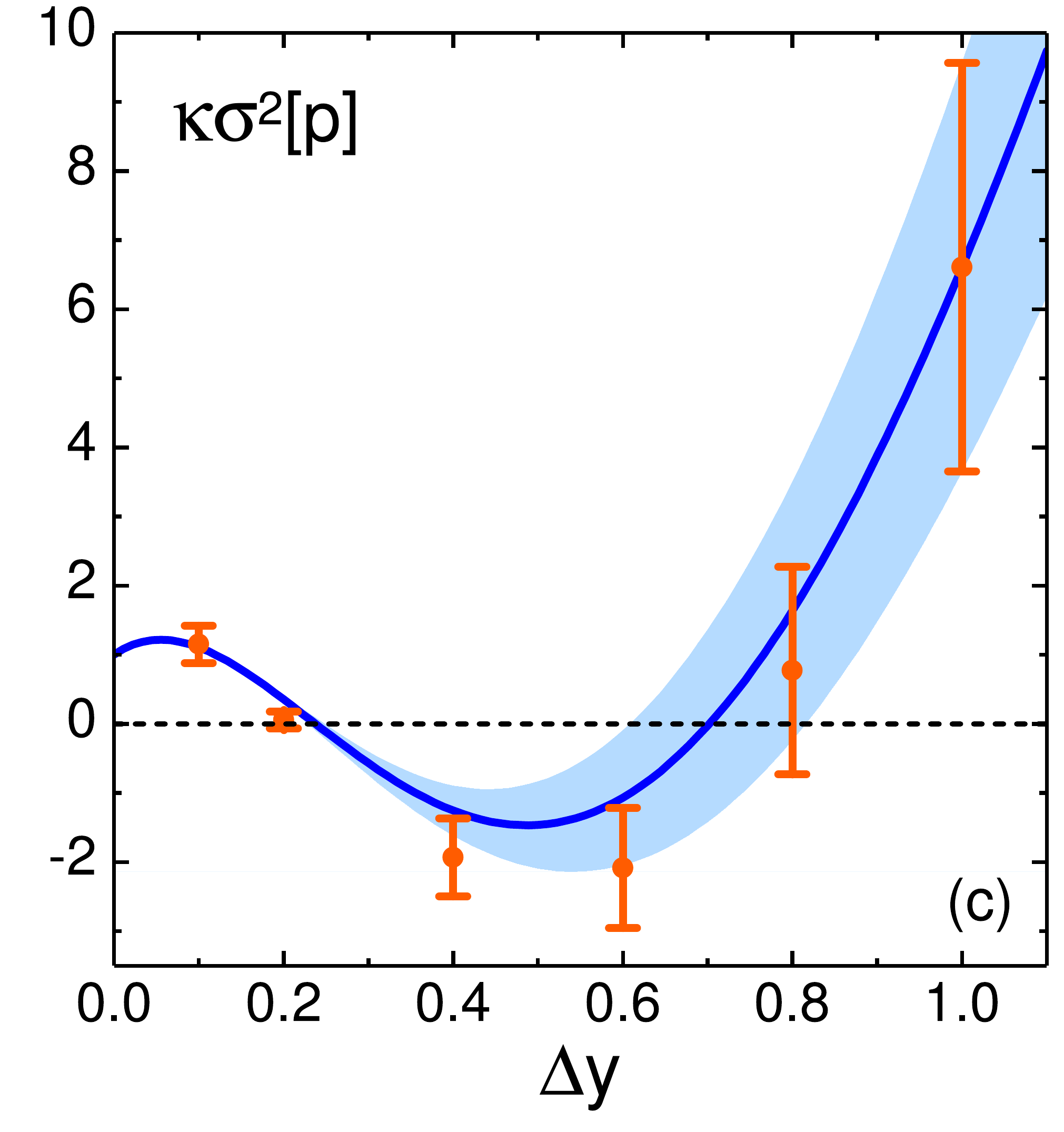}
\caption{Scaled variance (a), skewness (b), and kurtosis (c) of proton number fluctuations as functions of the rapidity interval $\Delta y$. The HADES data are shown by the symbols.  The  line corresponds to the binomial acceptance formulas~(\ref{w-alpha}-\ref{k-alpha}). The blue bands represent uncertainties due to HADES data errors in the $\Delta Y = 1$ rapidity interval.
}\label{fig-omega}
\end{figure*}

\section{ HADES results for proton number fluctuations}\label{HADES}
To describe the  proton number fluctuations in Au+Au central collisions  at $\sqrt{s_{\rm NN}}=2.42~{\rm GeV}$ as measured by the HADES Collaboration we use the 
binomial acceptance procedure
outlined in  Sec.~\ref{sec-binomial}. 
The HADES data for $\omega$, $S\sigma$, and $\kappa \sigma^2$ are presented for 6 symmetric rapidity intervals $\Delta y= 0.1,~0.2,~0.4~, 0.6,~0.8$, and $\Delta Y=1$ in the center of mass system. 
A transverse momentum cut $0.4 < p_T < 1.6$~GeV/$c$ was applied.
In what follows 
we view the largest rapidity interval $\Delta Y =1$ as a ``full" phase space. 
The quantities   $\omega[N]$, $S\sigma[N]$, and $\kappa \sigma^2[N]$ for this rapidity interval are then considered as values in the full space (\ref{omega}-\ref{kurt}). They are the input parameters 
for the binomial acceptance formulas (\ref{w-alpha}-\ref{k-alpha}).

The first step of the binomial acceptance procedure is calculating the corresponding $\alpha$-probabilities for different rapidity intervals.
In Fig.~\ref{fig} (a) the 
preliminary HADES data of the proton rapidity distribution for 10\% most central Au+Au collisions \cite{Szala:2020jro,PhysRevC.102.054903} are presented in the rapidity interval $\Delta Y=1$. 
We fit these data by the gaussian distribution 
\eq{
\frac{dN}{dy}~=~ 
C~\exp\left[ ~-\frac{y^2}{2a^2}\right] ~,\label{gauss}
}
with two parameters $C=90/\sqrt{2\pi a^2}$ 
and $a=0.62$ which estimate the height and the width of the distribution. For any $\Delta y \le 1$ one defines the $\alpha$-probabilities as
\eq{
\alpha~=~\frac{\int\limits_{- \Delta y/2} ^{\Delta y/2} \,dy~dN/dy} {\int\limits_{-1/2}^{1/2}\,dy
~dN/dy }~. 
\label{x}
}
The acceptance parameter $\alpha$ as a function of $\Delta y$
is shown in Fig.~\ref{fig} (b).

The scaled variance $\omega_{\alpha}[n]$, skewness $S\sigma_{\alpha}[n]$, and kurtosis $\kappa\sigma^2_{\alpha}[n]$ of proton number fluctuations as functions of $\Delta y$
are shown in Fig.~\ref{fig-omega}.
One can see that 
the binomial acceptance procedure gives a good agreement with the HADES data 
for all $\Delta y <1$. 
Thus, knowledge of ``global" cumulants 
(in the rapidity interval $\Delta Y=1$)
is sufficient to restore the corresponding values for any $\Delta y <1$, and no ``local" correlations between protons
within 
$\Delta Y$ are observed.

To find the signatures of these global fluctuations 
we suggest calculating
the correlation function for two arbitrary non-overlapping rapidity regions
$\Delta y_1$ and $\Delta y_2$, both inside the symmetric interval $\Delta Y=1$, 
with  $\langle N\rangle$ being the average number of protons  inside the interval $\Delta Y=1$  (see Appendix):
\eq{
\rho(n_1,n_2)\equiv ~ \langle N\rangle \frac{\langle n_1
n_2\rangle -\langle n_1\rangle \langle n_2\rangle}{\langle n_1\rangle \langle n_2\rangle} 
=\omega[N]-1~. \label{rho}
}
Equation (\ref{rho}) demonstrates the universal positive,
as $\omega[N]> 1$, 
correlations between $n_1$ and $n_2$. These correlations are independent of both the sizes of $\Delta y_1$ and $\Delta y_2$ and of their locations inside the rapidity interval $\Delta Y=1$. Note that for $\omega[N]= 0$ these correlations would be negative and equal $\rho=-1$  as a consequence of the global $N$-conservation in the $\Delta Y =1$ interval. The negative values  $-1 \le \rho \le 0$ correspond to small $N$-fluctuations with $0\le \omega[N]\le 1$.   
It would be interesting to check the relation (\ref{rho}) from the HADES data.

\section{Conclusions}\label{concl}

The binomial acceptance procedure describes
the scaled variance, skewness, and kurtosis of proton number fluctuations
measured recently by HADES Collaboration in 5\% central Au+Au collisions at $\sqrt{s_{\rm NN}}=2.42$~GeV in 
multiple rapidity intervals.  
Binomial acceptance formulas connect the observed large proton number fluctuations in the rapidity interval $\Delta Y=1$ with the observed proton number fluctuations.
This is consistent with the absence of 
local correlations between proton momenta inside the rapidity interval $\Delta Y=1$.  

The existing HADES data 
show large non-gaussian fluctuations of the number of  
protons within the rapidity interval $\Delta Y=1$.
These large fluctuations can be due to anomalies in the equation of state of matter
created in the collision which manifest themselves as local interproton correlations in the  coordinate space.
However, large fluctuations can also emerge due to some global external reasons which are valid even for a system of non-interacting particles.

An evident reason for the global proton number fluctuations could be 
event-by-event fluctuations in the number of nucleon participants. 
One should exclude this trivial source of event-by-event fluctuations. At small collision energies this is not an easy problem: there is no clear criterion to distinguish between the spectator and participant nucleons. 
The HADES data are corrected for volume fluctuations \cite{HADES:2020wpc}. However, additional studies in this direction would be helpful.

Another complication at the collision energy this low is the 
significant presence of light nuclear fragments in the final state.  
The existence of a large fraction of baryons in the form of nuclear fragments can generate large fluctuations of the number of bare protons. 
Finally, collective flows of baryons 
at low collision energies appear to be rather small which causes a problem to transfer the particle correlations from coordinate to momentum space.

An interesting consequence of the picture with the global proton number fluctuations and no local correlations between proton momenta is a universal form (\ref{rho})
for the correlations of multiplicities in two arbitrary non-overlapping rapidity intervals $\Delta y_1$ and $\Delta y_2$, both  inside the  rapidity region $\Delta Y=1$.
The relation (\ref{rho}) can be checked using the existing HADES data for protons at $\sqrt{s_{\rm NN}}=2.42~{\rm GeV}$.

	\begin{acknowledgments}
	The authors thank M. Gazdzicki, L.~Satarov, S.~Pratt, T.~Galatyuk, J.~Steinheimer, H.~Stoecker, and V.~Vovchenko for fruitful comments and discussions. This work is supported by the National Academy of Sciences of Ukraine, Grant No.  0122U200259. OS acknowledges the scholarship grant from the GET$\_$INvolved Programme of FAIR/GSI. R.V.P. acknowledges the scholarship of the President of Ukraine. M.I.G. and R.V.P. acknowledge the support from the Alexander von Humboldt Foundation. 
	\end{acknowledgments}
\appendix
\section{Derivation of Eq.~(\ref{rho}) 
}
Let $\Delta y_1$ and $\Delta y_2$ be the non-overlapping rapidity regions, both inside the interval $[-0.5,0.5]$, containing $n_1$ and $n_2$ particles, respectively. The  number of particles $N$ in the interval $[-0.5,0.5]$ is described by the probability distribution $P(N)$.
For uncorrelated particles 
the particle number  distribution  $P(n_1,n_2; N)$
can be presented in the following form:
\eq{\label{P1P2}
P(n_1,n_2;N)=
P(N)B(N,n_2|\alpha_2)B(N-n_2, n_1|\alpha_1)~,
}
where  $\alpha_1=\langle n_1\rangle/(\langle N\rangle -\langle n_2\rangle)$  and $\alpha_2=\langle n_2\rangle /\langle N\rangle$.
Using Eq.~\ref{P1P2} one then finds
\eq{
\langle n_1n_2\rangle &= \sum_N \sum_{n_1,n_2} n_1n_2 ~
P(n_1,n_2;N)~\nonumber\\&=~\mean{n_1}\mean{n_2}\left(1+\frac{\omega[N]-1}{N}\right)~. 
}

\bibliography{main.bib}
\end{document}